\documentclass[showpacs,aps,amsmath,amssymb]{revtex4}
\usepackage[latin1]{inputenc}
\usepackage{graphicx}
\usepackage{bm}
\usepackage{array}
\usepackage{braket}
\usepackage{microtype}
\usepackage{wasysym}
\renewcommand{\baselinestretch}{1.0}

\begin{document}

\renewcommand{\baselinestretch}{1.0}

\title{How a single particle modifies the physical reality of two distant others simultaneously: a quantum nonlocality and weak value study}

\author{Bertúlio de Lima Bernardo$^{1}$, Askery Canabarro$^{2}$, and Sérgio Azevedo$^{3}$}

\affiliation{$^{1}$Departamento de Física, Universidade Federal de Campina Grande, Caixa Postal 10071, 58109-970 Campina Grande, Paraíba, Brazil\\
$^{2}$Grupo de Física da Matéria Condensada, Núcleo de Ciências Exatas - NCex, Campus Arapiraca, Universidade Federal de Alagoas, 57309-005 Arapiraca-AL, Brazil\\
$^{3}$Departamento de Física, Universidade Federal da Paraíba, Caixa Postal 5008, 58059-900 João Pessoa, PB, Brazil}

\email{bertulio.fisica@gmail.com}

\begin{abstract}
The concept of wave-particle duality, which is a key element of quantum theory, has been remarkably found to manifest itself in several experimental realizations as in the famous double-slit experiment. In this specific case, a single particle seems to travel through two separated slits simultaneously. Nevertheless, it is never possible to measure it in both slits, which naturally appears as a manifestation of the collapse postulate. In this respect, one could as well ask if it is possible to ``perceive'' the presence of the particle at the two slits simultaneously, once its collapse could be avoided. In this article, we use the recently proposed entanglement mediation protocol to provide a positive answer to this question. It is shown that a photon which behaves like a wave, i.e., which seems to be present in two distant locations at the same time, can modify two existing physical realities in these locations. Calculations of the ``weak trace'' left by such photon also enforce the validity of the present argumentation.   
\end{abstract}

\maketitle


\section*{Introduction}

Since its inception, quantum mechanics has proclaimed many remarkable and counterintuitive behaviors of both light and matter. Perhaps, one of its most intriguing features is the wave-particle duality, concisely demonstrated in the double-slit experiment, which, as famously stated by Feynman, contains the only mystery of the theory \cite{feynman}. The experiment reflects the fact that when many quantum particles are sent one-by-one towards a mask with two slits to be detected by a screen placed on the far side of the mask, two complementary behaviors can be observed; the so-called wave and particle scenarios. The former appears whether no information about which slit the particle has passed through is known. In this case the particle density verified on the screen corresponds to an interference pattern, namely, a wave behavior. On the other hand, the latter manifests itself whenever one tries to obtain some information about the path of the particles. As a consequence, the interference pattern disappears, giving rise to a simple classical addition of the patterns created by the particles which passed though each slit separately - particle behavior.  

Trying to intuitively understand these behaviors brings into question two important cornerstones of quantum mechanics. The first is the superposition principle, which is evident if we comprehend that wave interference effects could never be observed if the particles have passed through only one of the slits. The second is the collapse postulate, which must be invoked to understand the fact that, despite the particles behave as if they pass through the two slits, it is impossible to detect them at both slits simultaneously, due to the instantaneous collapse of the wavefunction at the region around the slit in which the particle is detected \cite{dirac,shad}. Specifically, this last point can bring about another important question: could a single quantum particle manifests its presence simultaneously at the two slits if its collapse could be avoided? Under a slightly different perspective, an indication that a single particle could modify two separated physical realities at the same time would represent an important step towards an ultimate answer to this question.             

In this paper, we revisit the entanglement mediation protocol, which was recently proposed by one of the authors \cite{bert}, and use it as a framework to answer this question. It is argued that a single photon can modify two remote physical realities simultaneously, in the sense we described above. To this end, we analyze the behavior of a group of photons which are sent one-by-one through an apparatus with two possible paths and demonstrate, with basis on the outcomes of the entanglement mediation protocol and the weak traces that the photons leave, that they always pass through both paths simultaneously. Nevertheless, prior to this analysis, we demonstrate that it is possible to observe, after many runs of the experiment, that the physical realities of two other groups of photons, which are confined to regions adjacent to each of the possible paths, but distant from each other, are modified. Consequently, the reunion of these two ideas leads us to conclude that a single photon can be ``perceived'' at two distant places at the same time.

In the next section we outline the entanglement mediation protocol. In Sec. III we use some elements of Sec. II to demonstrate how the physical realities of two distant groups of photons can have their physical realities modified by the existence of an intermediate group of photons, and show that each photon of the intermediate group necessarily passes through two different paths at the same. Section IV is devoted to reinforce the fact that the photons of the intermediate group is, indeed, present in both paths by calculating their ``past'' by means of the weak trace they leave, according to a recent model proposed by Vaidman \cite{vaidman}. In Sec. V we present our conclusions.

\section*{Entanglement Mediation}

The entanglement mediation protocol as proposed in Ref.\cite{bert} consists in  three incident (indistinguishable) photons 1, 2 and 3  entering, respectively, three Mach-Zehnder-like apparatuses MZ$_{1}$, MZ$_{2}$ and MZ$_{3}$, simultaneously, as shown in Fig. 1. The whole experiment is composed by six 50:50 beam splitters BS$_{1}$,  BS$_{2}$, BS$_{3}$, BS$_{4}$, BS$_{5}$ and BS$_{6}$; two mirrors M$_{1}$ and M$_{2}$; and two photon detectors D$_{1}$ and D$_{2}$. After the primary devices (BS$_{1}$, BS$_{2}$ and BS$_{3}$), each apparatus MZ$_{(1,2,3)}$ prepares photon (1,2,3) in a coherent superposition of two possible paths $\ket{A}_{(1,2,3)}$ and $\ket{B}_{(1,2,3)}$. The central beam splitter BS$_{4}$ is shared by apparatuses MZ$_{1}$ and MZ$_{2}$, while BS$_{5}$ is shared by MZ$_{2}$ and MZ$_{3}$. With this configuration, if photons 1 and 2 arrive simultaneously at the opposite ports of BS$_{4}$, or photons 2 and 3 arrive simultaneously at BS$_{5}$, a two-photon interference effect takes place such that the photons always leave the respective beam splitter through the same output port \cite{irvine,hong,mandel}. At the last stage of the experiment, only photon 2 is allowed to interfere with itself due to the presence of BS$_{6}$ .      

The input state of the system, after the beam splitters BS$_{1}$, BS$_{2}$ and BS$_{3}$, is given by:
\begin{equation}
\label{1}
\ket{I} = \frac{1}{2 \sqrt{2}}(\ket{A}_{1}+i\ket{B}_{1})\otimes(\ket{A}_{2}+i\ket{B}_{2})\otimes(\ket{A}_{3}+i\ket{B}_{3}).
\end{equation}   
Here, we are considering that whenever a photon is reflected, it acquires a relative phase factor $i$ \cite{penrose}. Let us now neglect all the cases in which two photons leave one of the apparatuses MZ$_{1}$, MZ$_{2}$ and MZ$_{3}$ together. In other words, we will study only the situations in which, after passing through the secondary devices BS$_{4}$, BS$_{5}$, M$_{1}$ and M$_{2}$, no photon has passed from one of the apparatuses to another. Under this restriction, the terms $\ket{A}_{1}\ket{A}_{2}$ and $\ket{B}_{2}\ket{A}_{3}$ in Eq.~\eqref{1}, which evolve under the relations $\ket{A}_{1}\ket{A}_{2} \rightarrow \frac{i}{\sqrt{2}}(\ket{2A_{1}}+\ket{2A_{2}})$ and $\ket{B}_{2}\ket{A}_{3} \rightarrow \frac{i}{\sqrt{2}}(\ket{2B_{2}}+\ket{2A_{3}})$, as a result of the two-photon interference effect at the central beam splitters BS$_{4}$ and BS$_{5}$, must be excluded. Then, we eliminate these terms from Eq.~\eqref{1}, which yields:  
\begin{eqnarray}
\label{2}
\ket{I} &\rightarrow& \frac{1}{2 \sqrt{2}}( - \ket{A}_{1}\ket{B}_{2}\ket{B}_{3} + i\ket{B}_{1}\ket{A}_{2}\ket{A}_{3} \nonumber\\
&&- \ket{B}_{1}\ket{A}_{2}\ket{B}_{3} -i \ket{B}_{1}\ket{B}_{2}\ket{B}_{3}).
\end{eqnarray}   
 
Since we are not considering the possibilities of a photon to pass from one apparatus to another, after the secondary devices the term $\ket{A}_{1}\ket{B}_{2}\ket{B}_{3}$ acquires an overall phase factor $-i$ due to the three reflections that occur in the paths $A_{1}$, $B_{2}$ and $B_{3}$, and acquire a reduction in amplitude by a factor of 1/2, because of the possible transmissions of photons 1 and 2 at BS$_{4}$ and BS$_{5}$, respectively, which we are neglecting. Thus, we have the following evolution:
\begin{equation}
\label{3}
\ket{A}_{1}\ket{B}_{2}\ket{B}_{3} \rightarrow - \frac{i}{2}\ket{A}_{1}\ket{B}_{2}\ket{B}_{3}.
\end{equation}     
Similarly, the term $\ket{B}_{1}\ket{A}_{2}\ket{A}_{3}$ evolves under the relation
\begin{equation}
\label{4}
\ket{B}_{1}\ket{A}_{2}\ket{A}_{3} \rightarrow - \frac{i}{2}\ket{B}_{1}\ket{A}_{2}\ket{A}_{3}.
\end{equation} 
\begin{figure}[htb]
\begin{center}
\includegraphics[height=3.65in]{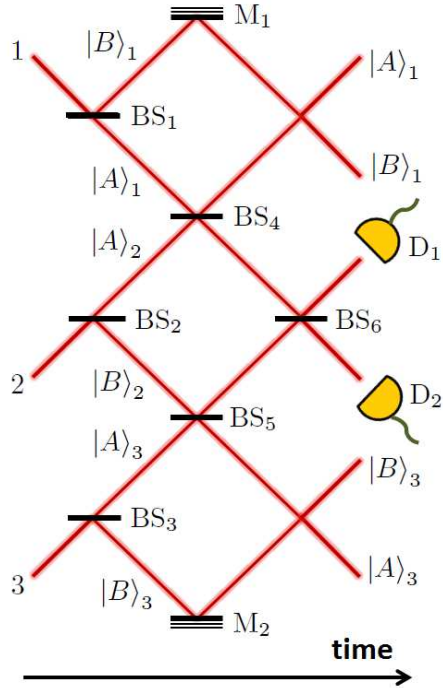}
\end{center}
\label{F2}
\caption{{\bf Experimental setup of the entanglement mediation protocol}. Three photons 1, 2 and 3 enter three Mach-Zehnder-like apparatuses, sharing two central beam splitters, BS$_{4}$ and BS$_{5}$. When we consider only the situations in which no transmission takes place at BS$_{4}$ and BS$_{5}$, after each run, depending upon whether D$_{1}$ or D$_{2}$ register photon 2, photons 1 and 3 end up, respectively, either in a non-maximally or a maximally entangled state. The direction of time in the laboratory reference frame is also indicated.}
\end{figure}
The $\ket{B}_{1}\ket{A}_{2}\ket{B}_{3}$ (and $\ket{B}_{1}\ket{B}_{2}\ket{B}_{3}$) term is also dictated by a similar evolution, but the amplitude now decreases by a factor of $1/\sqrt{2}$ due to the two reflections at M$_{1}$ and M$_{2}$ of the states $\ket{B}_{1}$ and $\ket{B}_{3}$, and only one possible transmission at BS$_{4}$ (and BS$_{5}$) because of the state $\ket{A_{2}}$ (and $\ket{B_{2}}$). Therefore, we have the evolutions
\begin{equation}
\label{5}
\ket{B}_{1}\ket{A}_{2}\ket{B}_{3} \rightarrow - \frac{i}{\sqrt{2}}\ket{B}_{1}\ket{A}_{2}\ket{B}_{3}
\end{equation}
and
\begin{equation}
\label{6}
\ket{B}_{1}\ket{B}_{2}\ket{B}_{3} \rightarrow - \frac{i}{\sqrt{2}}\ket{B}_{1}\ket{B}_{2}\ket{B}_{3}.
\end{equation}
By substitution of Eqs.~\eqref{3},~\eqref{4},~\eqref{5} and~\eqref{6} into Eq.~\eqref{2}, we find that the {\it normalized} state of the three photons after the secondary devices is given by   
\begin{eqnarray}
\label{7}
\ket{\psi} &=&  \frac{i}{\sqrt{6}} \ket{A}_{1}\ket{B}_{2}\ket{B}_{3} + \frac{1}{\sqrt{6}} \ket{B}_{1}\ket{A}_{2}\ket{A}_{3} \nonumber\\
&+& \frac{i}{\sqrt{3}}\ket{B}_{1}\ket{A}_{2}\ket{B}_{3} - \frac{1}{\sqrt{3}} \ket{B}_{1}\ket{B}_{2}\ket{B}_{3}.
\end{eqnarray}
The probability to obtain this state from the input state $\ket{I}$ of Eq.~\eqref{1} is given by $P = |\braket{\psi|I}|^{2} \approx 0.49$.

An important point to make clear is that, although the requirement of reflection of photons at the central beam splitters, the setup in Fig. 1 is fundamentally different from the situation in which the central beam splitters are replaced by mirrors. In fact, if we substitute BS$_{4}$ and BS$_{5}$ by mirrors, in virtue of the six possible reflections at the secondary devices, the evolution of the photons at this point of the experiment would be naturally given by $\ket{I} \rightarrow -i \ket{I}$, which is evidently different from the result of Eq.~\eqref{7}. The reason for this difference is that by positioning the central beam splitters rather than mirrors, and ignoring the cases in which two photons exit one of the apparatuses, one necessarily excludes all the possibilities that could cause the bunching effect, as performed in Eq.~\eqref{2}. In writing this equation, we eliminated the term $\ket{A}_{1}\ket{A}_{2}\ket{A}_{3}$ present in Eq.~\eqref{1}. Conversely, if there were mirrors instead of BS$_{4}$ and BS$_{5}$, this term would remain. A reasoning similar to the one that we used to obtain Eq.~\eqref{7} was developed in Ref. \cite{irvine}.

If we consider that the path lengths of MZ$_{2}$ are tuned so that if photon 2 is in the state $(i\ket{A}_{2}+\ket{B}_{2})/ \sqrt{2}$ before BS$_{6}$, then it will certainly exit from the lower port, i.e., the detector D$_{2}$ will certainly click. By the same token, if the state of photon 2 is  $(i\ket{A}_{2}-\ket{B}_{2})/ \sqrt{2}$ before BS$_{6}$, it certainly exits from the upper port, i.e., the detector D$_{1}$ certainly clicks. Now, let us investigate what happens to the state of photons 1 and 3 when photon 2 is detected at D$_{2}$. To this end, we project the state $\ket{\psi}$ onto the state $\ket{D}_{2} = (i\ket{A}_{2}+\ket{B}_{2})/ \sqrt{2}$ by means of the projector $\Pi_{2} = \ket{D}_{2} \bra{D}_{2}$. After the calculations, it can be found that the {\it normalized} resultant state of the system is
\begin{eqnarray}
\label{8}
\ket{n} = \frac{\Pi_{2}\ket{\psi}}{\sqrt{p(2)}}
&=& i \ket{D}_{2} \otimes \ket{\Psi^{-}}_{13},
\end{eqnarray}
with $p(2) = \braket{\psi|\Pi_{2}|\psi} = 1/6$ as the probability of detecting photon 2 at D$_{2}$ (See the note in \cite{note}). The result in Eq.~\eqref{8} is the tensor product of the state of photon 2 being detected at D$_{2}$ and a maximally entangled Bell state of photons 1 and 3, up to an overall phase factor $i$. Actually, once photon 2 is destroyed by detection, we are left with photons 1 and 3 in the singlet state:
\begin{eqnarray}
\label{9}
\ket{\Psi^{-}}_{13} = \frac{1}{\sqrt{2}} (\ket{A}_{1} \ket{B}_{3} - \ket{B}_{1} \ket{A}_{3}).
\end{eqnarray}
In part, it concludes the entanglement mediation protocol. In such a process one could maximally entangle photons 1 and 3 which were created in different sources and have never interacted. In the present case, as reported in Ref. \cite{bert}, photon 2, which exchanged information with photons 1 and 3 simultaneously at the two distant apparatuses MZ$_{1}$ and MZ$_{3}$, was the {\it mediator} of the entanglement.

On the other hand, in analyzing the state of the system when detector D$_{1}$ clicks, we apply the projector $\Pi_{1} = \ket{D}_{1} \bra{D}_{1}$ to the state $\ket{\psi}$, with $\ket{D}_{1} = (i\ket{A}_{2}-\ket{B}_{2})/ \sqrt{2}$, and obtain the following {\it normalized} state:   
\begin{eqnarray}
\label{10}
\ket{m} = \frac{\Pi_{1}\ket{\psi}}{\sqrt{p(1)}} &=&  \ket{D}_{1} \otimes \ket{\Phi}_{13},
\end{eqnarray}
with $p(1) = \braket{\psi|\Pi_{1}|\psi} = 5/6$ as the probability of detecting photon 2 at D$_{1}$ \cite{note}. The state $\ket{\Phi}_{13}$ of photons 1 and 3 is given by
\begin{equation}
\label{11}
\ket{\Phi}_{13} = \frac{1}{\sqrt{10}}(-i \ket{A}_{1}\ket{B}_{3} - i \ket{B}_{1}\ket{A}_{3} + 2 \sqrt{2} \ket{B}_{1}\ket{B}_{3}) ,
\end{equation}
which is a non-maximally entangled state with entanglement of formation given by E$(\Phi) \approx 0.08$ \cite{bert}. Note that, ultimately, a quantum correlation is established between photons 1 and 3, independently of measuring photon 2 at D$_{1}$ and D$_{2}$. It is important to make clear how the postselections into the states $\ket{n}$ and $\ket{m}$ can be performed. They are achieved simply by detecting a single photon at detector D$_{2}$ (and no photon at D$_{1}$) and a photon at D$_{1}$ (and no photon at D$_{2}$), respectively. Of course, the other two photons must be separately detected {\it a posteriori} at the outputs of MZ$_{1}$ and MZ$_{3}$.

Interestingly, it was also shown in Ref. \cite{bert} that the present arrangement brings about a manifestation of Hardy's paradox \cite{hardy,hardy2}, extended to three particles. Further, and more important, it leads us to conclude that photon 2, the mediator of the entanglement, can exchange information with both photons 1 and 3 simultaneously, which enforces the idea that a single quantum particle can pass through two different paths at the same time. In principle, the entanglement mediation protocol can be straightforwardly extended to massive particles. As a matter of fact, an atomic analog of the two-particle interference (Hong-Ou-Mandel effect), which supports the protocol, was recently experimentally realized  \cite{lopes}.    

\section*{Can the mediator modify two physical realities at the same time?}
           
In this section we want to analyze how photon 2 simultaneously influence the results of measurements made on photons 1 and 3 separately. Let us suppose that there exists an observer, ``Alice'', who is in charge of analyzing photon 1 in the apparatus MZ$_{1}$ (composed by the devices BS$_{1}$, M$_{1}$ and BS$_{4}$), localized at the output ports of it to measure the probabilities to obtain photon 1 in the paths A$_{1}$ and B$_{1}$, in accordance with Fig. 1; and another observer, ``Bob'', who is in charge of analyzing photon 3 in the apparatus MZ$_{3}$ (composed by the devices BS$_{3}$, M$_{2}$ and BS$_{5}$), localized at the outputs of it to measure the probabilities to obtain photon 3 in the paths A$_{3}$ and B$_{3}$. 

If Alice and Bob only know about the existence of photons 1 and 3 and their apparatuses MZ$_{1}$ and MZ$_{3}$, which results should they expect? To answer this question we have first to think about the apparatuses MZ$_{1}$ and MZ$_{3}$ individually, or, put differently, the complete apparatus of Fig. 1 in the absence of photon 2, given that photons 1 and 3 remain in their respective interferometers MZ$_{1}$ and MZ$_{3}$. In this case, we have that after passing through BS$_{1}$ the state of photon 1 is $(\ket{A}_{1} + i\ket{B}_{1})/\sqrt{2}$, which due to the reflection at M$_{1}$ and the equal probabilities of transmission (loss) and reflection at BS$_{4}$, evolves under the relations $\ket{B}_{1} \rightarrow i \ket{B}_{1}$ and $\ket{A}_{1} \rightarrow (i/\sqrt{2})\ket{A}_{1}$, respectively. Nevertheless, since we are interested in the cases in which Alice does measure photon 1 at one of the two outputs of MZ$_{1}$, the state of photon 1 after M$_{1}$ and B$S_{4}$ is given by the {\it normalized} state 
\begin{equation}
\label{12}
\ket{s}_{1} = (i\ket{A}_{1} - \sqrt{2}\ket{B}_{1})/\sqrt{3}.
\end{equation}
By a similar analysis, if we are interested in the cases in which Bob necessarily measures photon 3 at one of the output ports of MZ$_{3}$, the state of photon 3 after M$_{2}$ and B$S_{5}$ is found to be 
\begin{equation}
\label{12a}
\ket{s}_{3} = (i\ket{A}_{3} - \sqrt{2}\ket{B}_{3})/\sqrt{3}.
\end{equation}
In this case, the state of photons 1 and 3 after M$_{1}$, M$_{2}$, B$S_{4}$ and B$S_{5}$ is simply  
\begin{equation}
\label{12b}
\ket{\psi}_{13} = \ket{s}_{1} \ket{s}_{3} = \frac{1}{3}(i\ket{A}_{1} - \sqrt{2}\ket{B}_{1}) \otimes (i\ket{A}_{3} - \sqrt{2}\ket{B}_{3}).
\end{equation}
As a consequence, Alice must expect that the probabilities to obtain photon 1 in the states $\ket{A}_{1}$ and $\ket{B}_{1}$, given that photon 3 remains in MZ$_{3}$, are  respectively
\begin{equation}
\label{12c}
p(A_{1})= |(\bra{A}_{1} \bra{A}_{3}) \ket{\psi}_{13}|^{2} + |(\bra{A}_{1} \bra{B}_{3}) \ket{\psi}_{13}|^{2}= 1/3.
\end{equation}
and
\begin{equation}
\label{12d}
p(B_{1})= |(\bra{B}_{1} \bra{A}_{3}) \ket{\psi}_{13}|^{2} + |(\bra{B}_{1} \bra{B}_{3}) \ket{\psi}_{13}|^{2}= 2/3.
\end{equation}
In a similar fashion, Bob must expect that the probabilities to obtain photon 3 in the states $\ket{A}_{3}$ and $\ket{B}_{3}$, given that photon 1 remains in MZ$_{1}$, are  
\begin{equation}
\label{12e}
p(A_{3})= |(\bra{A}_{1} \bra{A}_{3}) \ket{\psi}_{13}|^{2} + |(\bra{B}_{1} \bra{A}_{3}) \ket{\psi}_{13}|^{2}= 1/3.
\end{equation}
and
\begin{equation}
\label{12f}
p(B_{3})= |(\bra{A}_{1} \bra{B}_{3}) \ket{\psi}_{13}|^{2} + |(\bra{B}_{1} \bra{B}_{3}) \ket{\psi}_{13}|^{2}= 2/3,
\end{equation}
respectively.

Let us now analyze the complete arrangement of Fig. 1 (including photon 2) in order to obtain the probabilities that Alice and Bob would, in fact, measure. In the case of Alice, the probability $P(A_{1})$ of obtaining, for example, A$_{1}$ as a result is given by the average of the probabilities to obtain A$_{1}$ when the states $\ket{m}$ and $\ket{n}$ are postselected, weighted over the probabilities, $p(1)$ and $p(2)$, to obtain these states. In this form, we have that   
\begin{equation}
\label{13}
P(A_{1})= p(1)|\bra{A}_{1}\ket{m}|^{2} + p(2)|\bra{A}_{1}\ket{n}|^{2} = 1/6.
\end{equation}
Analogously, the overall probability to obtain B$_{1}$ as a result is given by
\begin{equation}
\label{14}
P(B_{1})= p(1)|\bra{B}_{1}\ket{m}|^{2} + p(2)|\bra{B}_{1}\ket{n}|^{2} = 5/6.
\end{equation}
With similar calculations, it can also be found that the probabilities for Bob to find photon 3 in the paths A$_{3}$ and B$_{3}$ are given by $P(A_{3}) = 1/6$ and $P(B_{3}) = 5/6$, respectively. 

Now, we call attention to the fact that $P(A_{1}) \neq p(A_{1})$, $P(B_{1}) \neq p(B_{1})$, $P(A_{3}) \neq p(A_{3})$ and $P(B_{3}) \neq p(B_{3})$. It signifies that Alice and Bob will measure an unexpected result and conclude that some external influence must have disturbed their measurements. Undoubtedly, this external influence is a result of the presence of photon 2 in the intermediate apparatus MZ$_{2}$ (composed by the beam splitters BS$_{2}$, BS$_{4}$, BS$_{5}$ and BS$_{6}$), which was the only difference between the two cases presented above. In other words, the probabilities for Alice and Bob to detect photons 1 and 3 at the output ports of the apparatuses MZ$_{1}$ and MZ$_{3}$, respectively, depend on whether photon 2 is present in the setup of Fig. 1 or not. Then, we can conclude that photon 2 alone is able to modify the physical realities of two events (the detection probabilities of photons 1 and 3 at the output ports of the apparatuses MZ$_{1}$ and MZ$_{3}$). 

The discrepancy between the expected and measured results of the detection probabilities can only be realized by Alice and Bob after many runs of the experiment. Therefore, one could say that photon 2 disturbs either photon 1 or photon 3, one at a time, in the sense that when photon 2 goes through the path A$_{2}$ (B$_{2}$) it disturbs photon 1 (3), and that the discrepancy shown above takes place only as an average of the these many local disturbances. However, as shown in the previous section, and in Ref. \cite{bert}, we have two good reasons to argue that, with the configuration shown in Fig. 1, photon 2 always passes through both paths A$_{2}$ and B$_{2}$, instead of through only one at a time. First, because the probabilities of detecting photon 2 at detectors D$_{1}$ and D$_{2}$, which are given by p(1) = 5/6 and p(2) = 1/6, respectively, clearly indicate interference between the states $\ket{A_{2}}$ and $\ket{B_{2}}$ (in the absence of interference, one would expect p(1) = p(2) = 1/2). Second, and more important, in 100\% of the cases photon 2 is able to establish entanglement between photons 1 and 3, i.e., photon 2 always exchange information with photons 1 and 3 simultaneously \cite{bert}. Then, given that with the setup of Fig. 1, photons 1 and 3 have the probabilities of detection considerably disturbed by the presence of photon 2, and that photon 2 passes through both paths A$_{2}$ and B$_{2}$, the present scenario strongly suggests that photon 2 can modify the physical realities of both photons 1 and 3, which are distant and share no common past, simultaneously. This is the central result of this article.      

\section*{Asking about the past of the photons}

In general, a precise description about the past (or the future) of a particle is not clear in quantum theory. For example, we are not allowed to talk about the position of a quantum particle between two measurements, or its spin component along the $x$ direction immediately after the $z$ component has been measured. This is because a quantum measurement disturbs the system in a usually unpredictable way. However, a method for analyzing the past of a photon in a multipath interferometer has been recently proposed by L. Vaidman \cite{vaidman}, and experimentally tested by A. Danan {\it et. al.} \cite{danan}, which provided results that defy our common sense description about trajectories. 

Basically, the method consists in analyzing the {\it weak trace} left by the photon in all possible paths of an interferometer and, hence, use the results to argue where the photon has been prior to detection. The weak trace is assumed to be proportional to the weak value of the particle's projection operator in a given location, which, in turn, is the outcome of the weak measurement of the projective operator in question, which takes place between two strong measurements of the system, namely, the preselection and postselection \cite{vaidman,aharo,aha,kofman,dress}. The characteristic trait of a weak measurement is the fact that the perturbation of the state of the measured system is extremely small, such that the collapse of the wavefunction, which is an unavoidable feature in the standard (strong) measurement scheme \cite{neumann}, is prevented. This seems to be an ideal scenario to account for the problem that we are studying here. Nevertheless, the price to pay is that, in order to obtain the weak value, one must rely on weak measurements performed over a large ensemble of equally prepared systems before discussing the properties of each one individually. In other words, we must assume that the weak trace observed from the ensemble provides us the weak trace of each constituent particle \cite{vaidman}. A similar description of the past of a quantum particle was already used to study more than one particle at a time in the famous analysis of Hardy's paradox authored by Aharonov {\it et. al.} \cite{aharo2}, which was successfully demonstrated in recent experiments \cite{lundeen,yokota}. 

In what follows, we apply the weak trace method to verify where the three photons have passed through in the scheme of Fig. 1, when the states $\ket{\Phi}_{13}$ and $\ket{\Psi^{-}}_{13}$ are obtained as a result of the entanglement mediation protocol. In doing so, we will consider that the initial state of the system, the preselection, is given by the state $\ket{\psi}$, and the postselection will be either the state $\ket{n}$ of Eq.~\eqref{8}, or the state $\ket{m}$ of Eq.~\eqref{10}. Therefore, these two cases must be studied separately.

\subsection*{Weak traces of the photons when the state $\ket{n}$ is postselected}    

Let us investigate now the entanglement mediation protocol by using weak measurements when the singlet state $\ket{\Psi^{-}}_{13}$ of photons 1 and 3 is obtained after detecting photon 2 at D$_{2}$. We are interested in knowing where the photons are after they have passed through the secondary devices BS$_{4}$, BS$_{5}$, M$_{1}$ and M$_{2}$. As we have seen, the state of the system at this stage is $\ket{\psi}$, given in Eq.~\eqref{7}, which is the preselected state. Since we are now restricting ourselves to the study of the system when the state  $\ket{\Psi^{-}}_{13}$ of photons 1 and 3 is obtained, therefore, we consider $\ket{n}$ (the correspondent state of the three photons) as the postselected state in this subsection. In this scheme, the weak trace left, for example, by photon 1 along the path A$_{1}$ is given by the weak value of the projection operator $\pi_{A_{1}} = \ket{A}_{1}\bra{A}_{1}$ related to this path. Thus, we have that
\begin{equation}
\label{15}
\langle \pi^{(n)}_{A_{1}} \rangle_{w} = \frac{\braket{n|\pi_{A_{1}}|\psi}}{\braket{n|\psi}}= \frac{1}{2},
\end{equation}
where the superscript $n$ in the left-hand side of this equation denotes the postselected state to be considered. Similarly, it can be found that the weak traces for the other paths are given by
\begin{eqnarray}
\label{16}
\langle \pi^{(n)}_{B_{1}} \rangle_{w} = 1/2, &&\qquad \langle \pi^{(n)}_{A_{2}} \rangle_{w} = 1/2, \nonumber\\
\langle \pi^{(n)}_{B_{2}} \rangle_{w} = 1/2, &&\qquad \langle \pi^{(n)}_{A_{3}} \rangle_{w} = 1/2, \nonumber\\ \langle \pi^{(n)}_{B_{3}} \rangle_{w} = 1/2, &&
\end{eqnarray}
where $\pi_{B_{1}} = \ket{B}_{1}\bra{B}_{1}$,  $\pi_{A_{2}} = \ket{A}_{2}\bra{A}_{2}$, and so forth. According to Ref. \cite{vaidman}, the results of Eqs.~\eqref{15} and~\eqref{16} give us information about the past of each photon. As we can see, it tells us that the weak traces left by each photon along the two possible paths that they could pass through have the same magnitude. In other words, it can be said that photons 1, 2 and 3 passed through both the respective paths A and B simultaneously, leaving the same trace along them. In the case of photon 2, this result agrees with what we have presented in the previous section, in the sense that, since photon 2 always passes through both paths A$_{2}$ and B$_{2}$ at the same time, it alone could modify the physical realities of the distant photons 1 and 3 simultaneously.  

We call attention to the fact that Vaidman's theory on the past of a quantum particle is only applied to a single particle individually \cite{vaidman}, i.e., it says nothing about, for example, the simultaneous pasts of two (or three) photons inside a multipath arrangement like the one exhibited in Fig. 1. Even so, for the sake of completeness, we now calculate the (joint) weak values associated to the simultaneous traces left by photons 1, 2 and 3 in our scheme in order to obtain some new information. For this purpose, we define the following {\it joint} occupation operators  
\begin{eqnarray}
\label{17}
 \pi_{A_{1},A_{2},A_{3}} = \pi_{A_{1}}\pi_{A_{2}}\pi_{A_{3}}, &&\quad \pi_{A_{1},A_{2},B_{3}} = \pi_{A_{1}}\pi_{A_{2}}\pi_{B_{3}}, \nonumber\\
\pi_{A_{1},B_{2},A_{3}} = \pi_{A_{1}}\pi_{B_{2}}\pi_{A_{3}}, &&\quad \pi_{A_{1},B_{2},B_{3}} = \pi_{A_{1}}\pi_{B_{2}}\pi_{B_{3}}, \nonumber\\
\pi_{B_{1},A_{2},A_{3}} = \pi_{B_{1}}\pi_{A_{2}}\pi_{A_{3}}, &&\quad \pi_{B_{1},A_{2},B_{3}} = \pi_{B_{1}}\pi_{A_{2}}\pi_{B_{3}}, \nonumber\\
\pi_{B_{1},B_{2},A_{3}} = \pi_{B_{1}}\pi_{B_{2}}\pi_{A_{3}}, &&\quad \pi_{B_{1},B_{2},B_{3}} = \pi_{B_{1}}\pi_{B_{2}}\pi_{B_{3}}, \nonumber\\
\end{eqnarray}
which tell us about the simultaneous locations of photons 1, 2 and 3 in all possible paths. Before calculating the weak values of these operators, it is important to emphasize that the weak value of a product of operators is {\it not} equal to the product of the weak values of each operator individually \cite{resch2,kob,bert2}, i.e., the apparatus to obtain the weak traces of Eq.~\eqref{16} must be modified in order to measure the weak values of the operators of Eq.~\eqref{17}. In turn, after some calculations, the {\it joint} weak values can be found to be
\begin{eqnarray}
\label{18}
\langle \pi^{(n)}_{A_{1},B_{2},B_{3}} \rangle_{w} = 1/2, &&\quad  
\langle \pi^{(n)}_{B_{1},A_{2},A_{3}}\rangle_{w} = 1/2, 
\end{eqnarray}
and all the other joint weak values are zero. It signifies that the measurement apparatus only registers the simultaneous presence of photons 1, 2 and 3 in the paths (A$_{1}$,B$_{2}$,B$_{3}$) and (B$_{1}$,A$_{2}$,A$_{3}$). Note that these two paths do not allow the presence of two photons in the opposite sides of the central beam splitters BS$_{4}$ and BS$_{5}$. Indeed, if two photons could have been found simultaneously in the opposite sides of these beam splitters, the photon bunching effect would have taken place, and the state $\ket{\psi}$ could not have been preselected (See the discussion in Sec. II). It can be explained by the fact that the preselected state $\ket{\psi}$ of Eq.~\eqref{7} excludes the possibility of photon bunching at BS$_{4}$ and BS$_{5}$.

\subsection*{Weak traces of the photons when the state $\ket{m}$ is postselected}

Now we turn to the analysis of the weak traces left by photons 1, 2 and 3 when the state $\ket{\Phi}_{13}$ of photons 1 and 3 is postselected in the entanglement mediation protocol. In this respect, the correspondent state of the three photons is the state $\ket{m}$ given in Eq.~\eqref{10}. Similar to the case of the previous subsection, the traces left by the photons are given by the weak values of the projection operators, given that the state $\ket{\psi}$ is preselected and the state $\ket{m}$ is postselected. In doing so, for example, the weak trace left by photon 1 along the path A$_{1}$ is given by 
\begin{equation}
\label{19}
\langle \pi^{(m)}_{A_{1}} \rangle_{w} = \frac{\braket{m|\pi_{A_{1}}|\psi}}{\braket{m|\psi}}= \frac{1}{10}.
\end{equation}
Accordingly, the other weak traces can be found to be
\begin{eqnarray}
\label{20}
&&\langle \pi^{(m)}_{B_{1}} \rangle_{w} = 9/10, \qquad \; \; \langle \pi^{(m)}_{A_{2}} \rangle_{w} = 1/2, \nonumber\\
&&\langle \pi^{(m)}_{B_{2}} \rangle_{w} = 1/2, \quad \quad \quad \langle \pi^{(m)}_{A_{3}} \rangle_{w} = 1/10, \nonumber\\ &&\langle \pi^{(m)}_{B_{3}} \rangle_{w} = 9/10. 
\end{eqnarray}
Similar to the weak traces left when $\ket{n}$ is postselected, Eqs.~\eqref{15} and~\eqref{16}, these results tell us that each photon leaves a weak trace along the two paths through which they are allowed to pass. However, the traces left by photons 1 and 3 are asymmetric, in a manner that the magnitudes measured along the more external paths, B$_{1}$ and B$_{3}$, are nine times bigger than those along the more central ones, A$_{1}$ and A$_{3}$. In the case of photon 2, we find again that it leaves traces of equal magnitude in both paths A$_{2}$ and B$_{2}$, which also enforces the idea that this photon passed through these two paths at the same time. This particularity was observed with the two possible postselections $\ket{n}$ and $\ket{m}$.

Finally, we verify the weak values of the joint occupation operators given in Eq.~\eqref{17} when the state $\ket{m}$ is postselected. In this case, it can be found that    
\begin{eqnarray}
\label{21}
&& \langle \pi^{(m)}_{A_{1},B_{2},B_{3}} \rangle_{w} = 1/10,   \quad  
\langle \pi^{(m)}_{B_{1},A_{2},A_{3}}\rangle_{w} = 1/10, \nonumber\\ 
&& \langle \pi^{(m)}_{B_{1},A_{2},B_{3}} \rangle_{w} = 2/5,  \  \; \quad  
\langle \pi^{(m)}_{B_{1},B_{2},B_{3}}\rangle_{w} = 2/5,
\end{eqnarray}
and that all the other joint weak values are zero. Again, we observe that the joint weak values are non-vanishing for the paths in which two photons cannot be found in the opposite sides of the central beam splitters BS$_{4}$ and BS$_{5}$.  Nonetheless, we observe now some asymmetry among the magnitudes of the calculated weak values. The paths (B$_{1}$,A$_{2}$,B$_{3}$) and (B$_{1}$,B$_{2}$,B$_{3}$), which present vanishing joint weak values when $\ket{n}$ is postselected, exhibit non-vanishing weak values in the present case.   

\subsection*{Connection between weak values and expectation values}

An important property of weak measurements is that, given a general quantum state $\ket{\psi}$, one can define the expectation value of an observable, $\langle \hat{A} \rangle$, by expanding it in terms of a complete set of ket states $\ket{\phi_{n}}$ in a specific form \cite{aharo3}:
\begin{eqnarray}
\label{22}
\langle \hat{A} \rangle &=& \braket{\psi|\hat{A}|\psi} = \bra{\psi} \sum_{n} \ket{\phi_{n}}\braket{\phi_{n}|\hat{A}|\psi}
 \nonumber\\ &=& \sum_{n} |\braket{\psi|\phi_{n}}|^{2} \frac{\braket{\phi_{n}|\hat{A}|\psi}}{\braket{\phi_{n}|\psi}} = \sum_{n} P(n) \langle \hat{A} \rangle_{w} (n), \nonumber\\  
\end{eqnarray}    
where $P(n)$ is the probability to obtain the state $\ket{\phi_{n}}$ from the state $\ket{\psi}$, and $\langle \hat{A} \rangle_{w} (n)$ is the weak value of $\hat{A}$ when the sate $\ket{\psi}$ is preselected and $\ket{\phi_{n}}$ is postselected.

In regards to our problem, we can estimate the probabilities to measure photon 1 in the states $\ket{A_{1}}$ and $\ket{B_{1}}$, and photon 3 in the states $\ket{A_{3}}$ and $\ket{B_{3}}$. Naturally, these probabilities can be calculated by means of the expectation value of the projection operators in each of these states. Thus, by using the result of Eq.~\eqref{22}, taking into account the two possible postselections $\ket{m}$ and $\ket{n}$, we have that     
\begin{equation}
\label{23}
P(A_{1})= \langle \pi_{A_{1}} \rangle=  p(1) \langle \pi^{(m)}_{A_{1}} \rangle_{w} + p(2) \langle \pi^{(n)}_{A_{1}} \rangle_{w} = 1/6.
\end{equation}
and 
\begin{equation}
\label{24}
P(B_{1})= \langle \pi_{B_{1}} \rangle=  p(1) \langle \pi^{(m)}_{B_{1}} \rangle_{w} + p(2) \langle \pi^{(n)}_{B_{1}} \rangle_{w} = 5/6.
\end{equation}
These results are in agreement with the probabilities found in the previous section, Eqs.~\eqref{13} and~\eqref{14}. Similarly, we can use this method to find that $P(A_{3}) = 1/6$ and $P(B_{3}) = 5/6$, which are also in agreement with the results of the previous section. In this form, the weak trace method also confirms that photon 2 alone can influence the outcomes of measurements made on photons 1 and 3 simultaneously, which concludes our analysis.

\section*{Conclusion}

In conclusion, we revisited the entanglement mediation protocol and, based on the fact that the presence of one photon, the mediator, can disturb the detection probabilities of two distant others, and that this mediator always passes through two paths simultaneously, we argue that it can modify two remote (and independent) physical realities at the same time. To this end, we first showed how the detection probabilities of the two separated photons, which share no common past, is disturbed by the presence of the mediator. Second, we assert that it is always present simultaneously in two distant locations because it interferes with itself and invariably mediates entanglement between the other photons in every run of the protocol. At last, we used the recently proposed weak trace method, which describes the past of a single particle, to confirm the simultaneous presence of the mediator at two distant locations of the setup. In general terms, when we analyze the consequences obtained from the present proposal and those from the double-slit experiment, we can see that the last case leads us to conclude that, since an interference pattern is created, each particle must have passed through both slits simultaneously, but whenever one tries to measure their paths, the wavefunction collapses and the interference pattern is destroyed, i.e., the presence of each particle is manifested uniquely at one of the possible paths. Conversely, in this work we took a step forward by showing that it is possible to make the mediator particle (photon 2) interfere with itself, at the same time that it exerts a physical and measurable influence on two distant photons simultaneously. In principle, these findings can be straightforwardly generalized to massive particles.     
        
\section*{Acknowledgements}
 
The authors acknowledge financial support from the Brazilian funding agency CNPq. A.C also thank PROPEP/UFAL, and the Alagoas State Research Agency FAPEAL through 
major projects (PPP - 20110902-011-0025-0069 / 60030-733/2011).

\section*{Author Contributions}
 
B.L.B contributed to the conception and development of the ideas. B.L.B, A.C, and S.A contributed to the writing of the manuscript and discussion of the results.  

\section*{Competing financial interests}

The authors declare no competing financial interests



\begin{thebibliography}{7}
\bibitem{note} The overall probability (with respect to the input state $\ket{I}$) is obtained by multiplying this result by the factor $P \approx 0.49$.
\end{thebibliography}
\end{document}